\documentstyle[12pt,aaspp4]{article}

\begin{document}

\title
{The $K$--band Luminosity Function in Galaxy Clusters to $z \sim 1$}

\author
{Roberto De Propris}

\affil
{School of Physics, Department of Astronomy \& Optics,
 University of New South Wales, Sydney, Australia, NSW 2052;
 propris@edwin.phys.unsw.edu.au }

\author
{S. A. Stanford\altaffilmark{1}}

\affil
{University of California at Davis, and the Institute of Geophysics and Planetary Physics,
 Lawrence Livermore National Laboratories,
 Livermore, CA, 94550; adam@igpp.llnl.gov}

\author
{Peter R. Eisenhardt\altaffilmark{1}}

\affil
{MS 169-327, Jet Propulsion Laboratory,
 California Institute of Technology,
 4800 Oak Grove Drive, Pasadena, CA, 91109;
 prme@kromos.jpl.nasa.gov}

\author
{Mark Dickinson\altaffilmark{1}}

\affil
{Space Telescope Science Institute, Baltimore, MD, 21218;
med@stsci.edu}

\author{Richard Elston\altaffilmark{1}}

\affil{Department of Astronomy, University of Florida,
       Gainesville, FL, 32611; elston@astro.ufl.edu}

\altaffiltext{1}{Visiting Astronomer, Kitt Peak National Observatory, National Optical Astronomy
Observatories, which is operated by the Association of Universities for Research in Astronomy,
Inc. (AURA) under cooperative agreement with the National Science Foundation.}

\begin{abstract}

We present $K$--band luminosity functions for galaxies in a
heterogeneous sample of 38 clusters at $0.1 < z < 1$.  Using
infrared--selected galaxy samples which generally reach 2 magnitudes
fainter than the characteristic galaxy luminosity $L^*$, we fit
Schechter functions to background--corrected cluster galaxy counts to
determine $K^*$ as a function of redshift.  Because of the magnitude
limit of our data, the faint-end slope $\alpha$ is fixed at $-0.9$ in
the fitting process.  We find that $K^*(z)$ departs from no--evolution
predictions at $z > 0.4$, and is consistent with the behavior of a
simple, passive luminosity evolution model in which galaxies form all
their stars in a single burst at $z_f = 2 (3)$ in an $H_0 = 65$ km
s$^{-1}$ Mpc$^{-1}$, $\Omega_M = 0.3$, $\Omega_{\Lambda}=0.7 (0)$
universe.  This differs from the flat or negative infrared luminosity
evolution which has been reported for high redshift {\it field} galaxy
samples.  We find that the observed evolution appears to be
insensitive to cluster X-ray luminosity or optical richness, implying
little variation in the evolutionary history of galaxies over the
range of environmental densities spanned by our cluster sample.  These
results support and extend previous analyses based on the color
evolution of high redshift cluster E/S0 galaxies, indicating not only
that their stellar populations formed at high redshift, but that the
assembly of the galaxies themselves was largely complete by $z \approx
1$, and that subsequent evolution down to the present epoch was
primarily passive.

\end{abstract}

\section
{Introduction}

The predominance of early--type galaxies in clusters compared to the 
field is a clear sign that the evolution of galaxies depends on environment, 
but the physical mechanisms at work remain the subject of considerable 
debate.  In particular, the question of when and how the most massive 
galaxies formed, and especially the origin of giant elliptical galaxies, 
is a topic of ongoing, vigorous theoretical and observational research.  
The traditional picture of elliptical galaxy formation (e.g., Eggen, 
Lynden-Bell \& Sandage 1962) postulates a single burst of star formation 
at high redshift followed by passive evolution.  The existence of a tight 
color--magnitude relation in nearby clusters (e.g., Bower, Lucey \& Ellis 1992; 
De Propris et al.\ 1999) and the close correlation between galaxy mass
and metallicity implied by the Mg$_2-\sigma$ relation (Bender,
Burstein \& Faber 1993) is explained naturally by a single, early 
episode of star formation.  If more massive galaxies are more efficient 
at retaining supernova ejecta, they will have higher metallicities and 
therefore redder colors (Arimoto \& Yoshii 1987).  If the color--magnitude 
relation is driven primarily by metal abundance, then the scatter in galaxy 
colors around the mean locus provides constraints on the epoch of galaxy 
formation.  Both Bower et al.\ (1992) and De Propris et al.\ (1999) conclude 
that the intrinsic scatter in the $U-V$ colors of Coma galaxies is consistent 
with high formation redshifts and/or extreme synchronization in the epoch of 
galaxy formation.

The picture outlined above may be overly simplistic.  A population of
blue galaxies is commonly encountered in clusters at moderate redshift,
with almost no counterpart in present day systems (Butcher \& Oemler
1984).  High resolution imaging indicates that these blue galaxies are
mostly late--type spirals undergoing starbursts (Dressler et al.\ 1994;
Couch et al.\ 1998 and references therein), and suggests that the 
cluster S0 population is far less abundant at $z \sim 0.5$ than today
(Dressler et al.\ 1997).  It has been proposed that the blue galaxies
are the progenitors of cluster S0 galaxies (e.g.\ Couch \& Sharples
1987).  In agreement with this explanation of the Butcher-Oemler
effect, some S0s in the present epoch appear to contain younger stellar
populations (Bothun \& Gregg 1990; Kuntschner \& Davies 1998).
However, the colors of the combined E and S0 populations in the core 
of clusters show little change, other than that expected from passive 
spectral evolution, out to $z \sim 1$, even among clusters of widely 
varying richness and X-ray luminosity (e.g., Stanford, Eisenhardt, 
\& Dickinson 1998, hereinafter SED98).  

Further complications in the monolithic collapse scenario arise if, as
has recently been found in the Coma cluster by J{\o}rgensen (1999), the
mean ages and abundances of cluster early-types are anti-correlated.
A relation between age and abundance could allow significant
variations in the ages of the stellar populations in cluster
early-type galaxies, while keeping the scatter in e.g.\ the
color-magnitude relations small.  A possible correlation between age
and metallicity has been emphasized previously by Worthey, Trager, \&
Faber (1996), wherein the oldest galaxies are the most metal-poor.
But recent work on the Fundamental Plane has found support for a
scenario in which age varies $directly$ with metallicity in elliptical
galaxies, i.e.\ the most luminous ellipticals are the oldest and most
metal rich, in a wide range of environments from the field to the
Coma cluster (Pahre, De Carvalho, \& Djorgovski 1998), suggesting
that there still exists considerable controversy in such studies.

Hierarchical structure formation models present a very different view
of galaxy evolution, in which galaxies assemble by a process of
gradually merging smaller stellar systems over a wide range of
redshifts (e.g., Cole et al.\ 1994).  Some key factors governing the
spectrophotometric evolution of elliptical galaxies in this scenario
are the time at which the bulk of the stellar populations are formed,
the era when the majority of mergers take place, and the amount of new
star formation induced during each merging event in the construction
of a large galaxy.  If the formation of a large fraction of the stars
in a giant elliptical galaxy is distributed over a broad cosmic time
interval, then ellipticals at any redshift should exhibit a wide range
in their mass--weighted stellar ages.  If, however, most stars in
present--day cluster ellipticals were formed in smaller disk galaxies
at large lookback times ($z \gg 1$), and if little additional star
formation takes place during subsequent mergers, then the end--product
ellipticals could appear to be old and approximately coeval even if
the bulk of the merging took place relatively late.  The hierarchical
models of Kauffmann \& Charlot (1998a) reproduce the color--magnitude
relation for ellipticals because more massive galaxies are the product
of systematically more massive progenitors which retain more metals
when forming their stars.

Discriminating between the formation scenarios described above is
difficult, even by studying galaxies in clusters at high redshift.
The evolution of the color magnitude relation to $z
\sim 1$ has been discussed by many workers in the field, e.g.\
Arag\'on-Salamanca et al.\ 1993, Lubin 1996, Ellis et al.\ 1997, and
SED98.  Most results are consistent with the high galaxy formation
redshifts and subsequent passive evolution favored by monolithic
collapse.  But even the most comprehensive study spanning the largest
redshift range, SED98, could not exclude the possibility that cluster
ellipticals formed more recently from mergers of smaller galaxies, as
long as the bulk of the stellar mass was created at much larger
redshifts and there was little recent star formation in the subsequent
merging process.  Kauffmann \& Charlot (1998a) have also argued that
by selecting galaxies exclusively from rich clusters, studies such as
those of Ellis et al.\ (1997) and SED98 introduce an increasingly
strong bias at higher redshifts toward galaxies that have formed at
earlier epochs, and therefore that the conclusions cannot be extended
to the general cluster galaxy population today.  An additional concern
is that SED98 and Ellis et al.\ 1997 selected galaxies only from the
densest core regions of clusters imaged by the {\it Hubble Space
Telescope (HST).}

An alternative and potentially more powerful means of testing models
of galaxy formation is to determine the mass function of galaxies over
a large redshift range.  The actual mass function is difficult to
measure, although recent kinematic studies of the fundamental plane of
cluster galaxies at high redshift (e.g.\ Kelson et al.\ 1997; van
Dokkum et al.\ 1998) have made an important step in this direction.
The $K$--band luminosity function (LF) may serve as a useful
surrogate.  Traditional optical LFs have been used as a sensitive
probe of the bulk properties of galaxy populations both in the local
universe (e.g., Binggeli, Sandage \& Tammann 1988) and at high
redshift (e.g.\ Lilly et al.\ 1995b; Ellis et al.\ 1996; Cowie et al.\
1996).  Although they are special environments, galaxy clusters are
convenient to study because their luminosity functions can be measured
without extensive spectroscopy by using statistical field galaxy
subtraction.  Infrared luminosities are particularly well suited for
defining luminosity functions because they broadly reflect the total
stellar mass of the galaxies, and do not depend strongly on the
details of their stellar populations (Gavazzi, Pierini, \& Boselli\
1996).  This allows us to study luminosity evolution at wavelengths
where the mass--to--light ratio is comparatively insensitive to the
star formation history (cf.\ Madau, Pozzetti, \& Dickinson 1998).  In
addition, (i) the effects of extinction are much weaker at infrared
wavelengths than in the optical; (ii) the infrared LF appears to be
independent of environment (De Propris et al.\ 1998), and (iii)
$k$--corrections for infrared colors are only weakly dependent on
Hubble type and vary slowly with redshift (e.g., Poggianti 1997).

Luminosities derived from observed magnitudes depend on the assumed
cosmology.  In addition, cosmology affects the evolution of galaxy
luminosities and colors by setting the relationship between lookback
time and redshift.  Jointly constraining the cosmological parameters,
galaxy formation redshifts, and star formation histories is beyond the
scope of this paper.  Therefore when comparing our data to
evolutionary models we will constrain the discussion to a limited set
of cosmologies.  We adopt the parameters $\Omega_M=0.3$ and
$\Omega_{\Lambda} =0.7$, with $H_0=65$ km s$^{-1}$ Mpc$^{-1}$ for
consistency (Perlmutter et al.\ 1998; Riess et al.\ 1998), as our
default cosmology.  These high redshift supernovae results currently
provide the strongest evidence for non-zero $\Lambda$ cosmologies.  In
our analysis, we also consider more traditional $\Omega_M=0.3$,
$\Omega_{\Lambda}=0.0$, and $\Omega_M=1$, $\Omega_{\Lambda}=0.0$
models.

In this paper we derive $K$--band LFs for a large sample of clusters
at $0.1 \lesssim z \lesssim 1$.  Since we remove field galaxy
contamination statistically, we are sensitive to all morphological
types and are not biased towards E/S0 galaxies over the covered
magnitude range, unlike SED98 where the spiral and irregular galaxies
were excluded.  Our sample consists of the clusters reported in SED98,
plus an additional set described below.  We use the entire fields
surveyed by our ground-based $K$--band imaging, rather than the
smaller {\it HST} fields studied by SED98.  The sample of 38 clusters
spans a large range in several cluster properties and so may
constitute a reasonably fair sample of cluster galaxy populations
brighter than $K^*+2$ over roughly half the age of the universe.

The structure of this paper is as follows.  In section 2 we describe
our observations and photometry.  Section 3 discusses star--galaxy
separation, statistical subtraction of field galaxies, and the
derivation of the cluster luminosity functions.  In section 4 we
analyze the results and consider their implications for cluster galaxy
evolution.

\section{Observations, Data Reduction, and Photometry}

\begin{figure}
\caption{Redshift distribution of the clusters used in the luminosity function analysis.}
\label{zhist}
\plotone{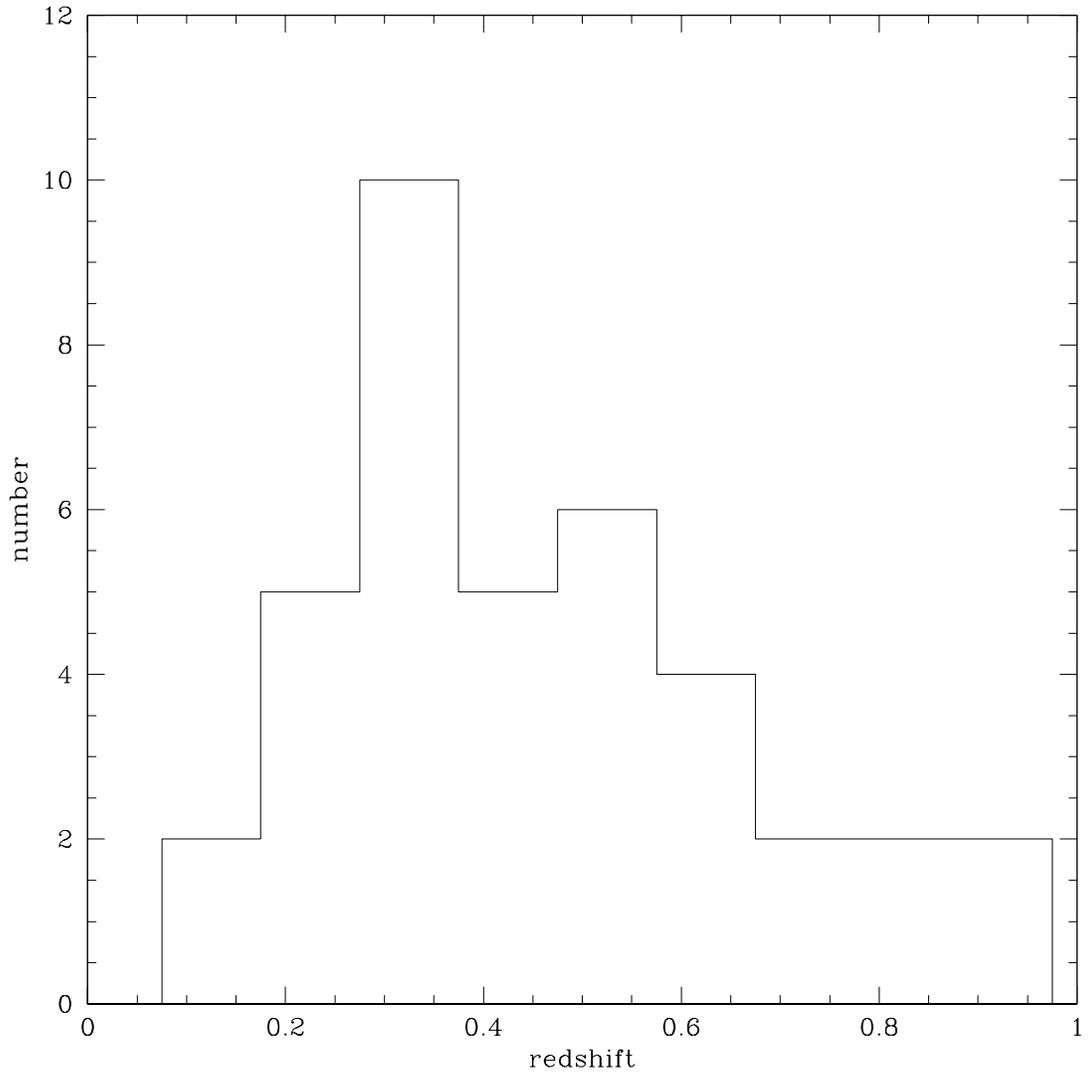}
\end{figure}

The overall sample of clusters presented here is large and
heterogeneous, consisting of 38 clusters at $0.14 < z < 0.92$, drawn
from a variety of optical, X-ray, and radio selected samples.  The
redshift distribution of the whole sample is shown in
Figure~\ref{zhist}.  As part of a survey more fully described in
SED98, $JHK_s$ images of the clusters were obtained using IR array
cameras on NOAO telescopes at Kitt Peak and Cerro Tololo in
1993--1996.  The lower redshift clusters were observed at the 2.1~m
telescope at KPNO using IRIM, which has a 256$\times$256 HgCdTe array
with 1.09 arcsec pixels, and at the 1.5~m telescope at CTIO using
CIRIM, which has a $256 \times 256$ HgCdTe array with 1.16 arcsec
pixels, and OSIRIS, which had a $256 \times 256$ HgCdTe array with
0.95 arcsec pixels.  The highest redshift clusters were observed at
the 4~m telescope at KPNO using IRIM, where it has 0.6 arcsec pixels.
Two of the clusters, Abell 370 and 851, were observed in 1991 with
SQIID, which had $256 \times 256$ PtSi detectors with 1.30 arcsec
pixels, as reported in Stanford, Eisenhardt \& Dickinson 1995
(hereinafter SED95).  The sample of clusters is summarized in
Table~\ref{samp}, along with details on the observations such as the
telescope/instrument, field size, and number of objects within the
field for each cluster.

Exposure times in all bandpasses were chosen to provide galaxy
photometry with $S/N > 5$ for galaxies with the spectral energy
distributions of present--day ellipticals down to $\sim$2 magnitudes
fainter than $L^\ast$ (unevolved) at the cluster redshift.  This
permits us to study galaxy properties over a similar range of
luminosities for all clusters in our sample, regardless of their
redshifts. Table 1 lists the $K$ magnitude limit used to define the
object sample for each cluster.  Our images typically cover a field
size of $\sim$1 to 1.8 Mpc at the cluster redshift; the actual field
size is given in Table 1 for each cluster.  The IR images were
calibrated onto the CIT system wherein Vega has $m = 0$.  The typical
rms of the transformations is 0.03.  The effective angular resolution
of the images is generally limited by the large pixel scale of the
infrared arrays, and is $\sim$1.7 arcsec for the $z < 0.6$ clusters
and $\sim$1.2 arcsec for the more distant objects.

Object detection was carried out on the $K$ images using a modified
version (Adelberger, personal communication) of FOCAS (Valdes 1982),
which was also used to obtain ``total'' magnitudes.  The observing
methods, data reduction techniques, and photometric methods for our
ground--based data set are described in more detail in SED95.  All
photometry has been corrected for reddening using the interstellar
extinction curve given in Mathis (1990), with values for $E(B-V)$
taken from Burstein \& Heiles (1982).

\section{Analysis}

In order to derive LFs for our clusters, we need to remove
contamination by stars in our Galaxy, and by background and foreground
galaxies.  Finally, we fit Schechter (1976) functions to the summed LF
of the clusters in discrete redshift bins.  We describe the procedures
employed in the next three subsections.

\subsection{Star-Galaxy Separation}

The $K$--band images generally lack sufficiently high angular resolution
to effectively distinguish stars from galaxies on the basis of image
concentration or sharpness.  Instead, we use color criteria and
models.  Huang et al.\ (1997) find that a line corresponding to the
equation $(B-I) - 2.5\,(I-K) > -2$ separates stars and galaxies
effectively in a color--color plane.  Our cluster data do not always
span such a large wavelength range; therefore we devise a different
criterion based on infrared colors, which are available for all our
objects.

\begin{figure}
\caption{ Color--magnitude diagram for objects in a subset of the
clusters with {\it HST} imaging.  Filled squares are stars, and open
circles are galaxies, as determined from archival WFPC2 images.  A $J-K=1$
line effectively separates stars and galaxies.}
\label{jkhst}
\epsscale{0.8}
\plotone{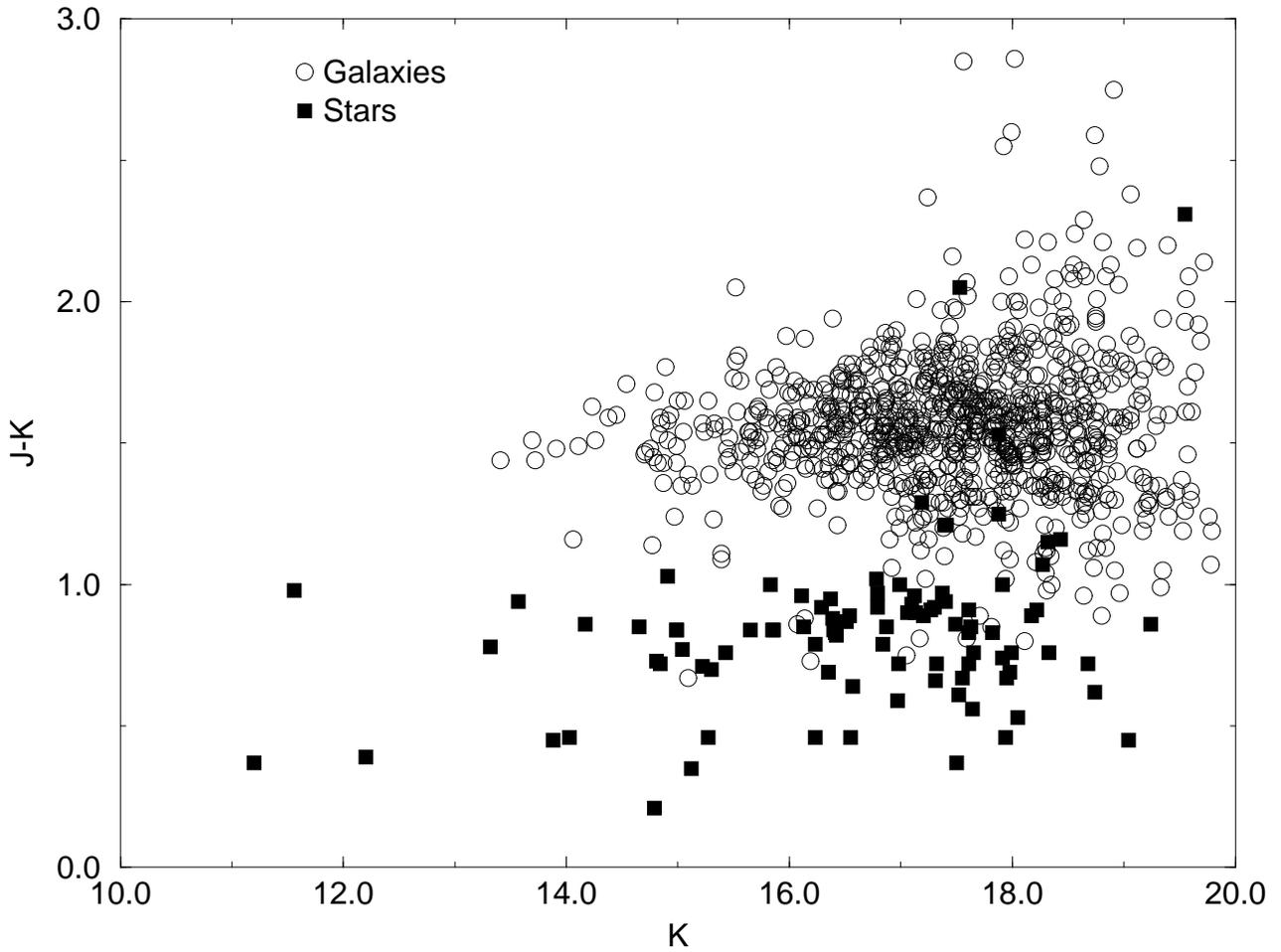}
\epsscale{1.0}
\end{figure}

\begin{figure}
\epsscale{0.7}
\caption{Star counts in the four $\sim$25 arcmin$^2$ fields observed 
fby Elston et al.\ (1999), selected using the color criterion
of Huang et al.\ 1997 (open squares),  and with our $J-K$ criterion 
(filled circles).  The thick solid lines show predictions from the
the models of Bahcall \& Soneira (1980, 1981).}
\label{jkstars}
\plotone{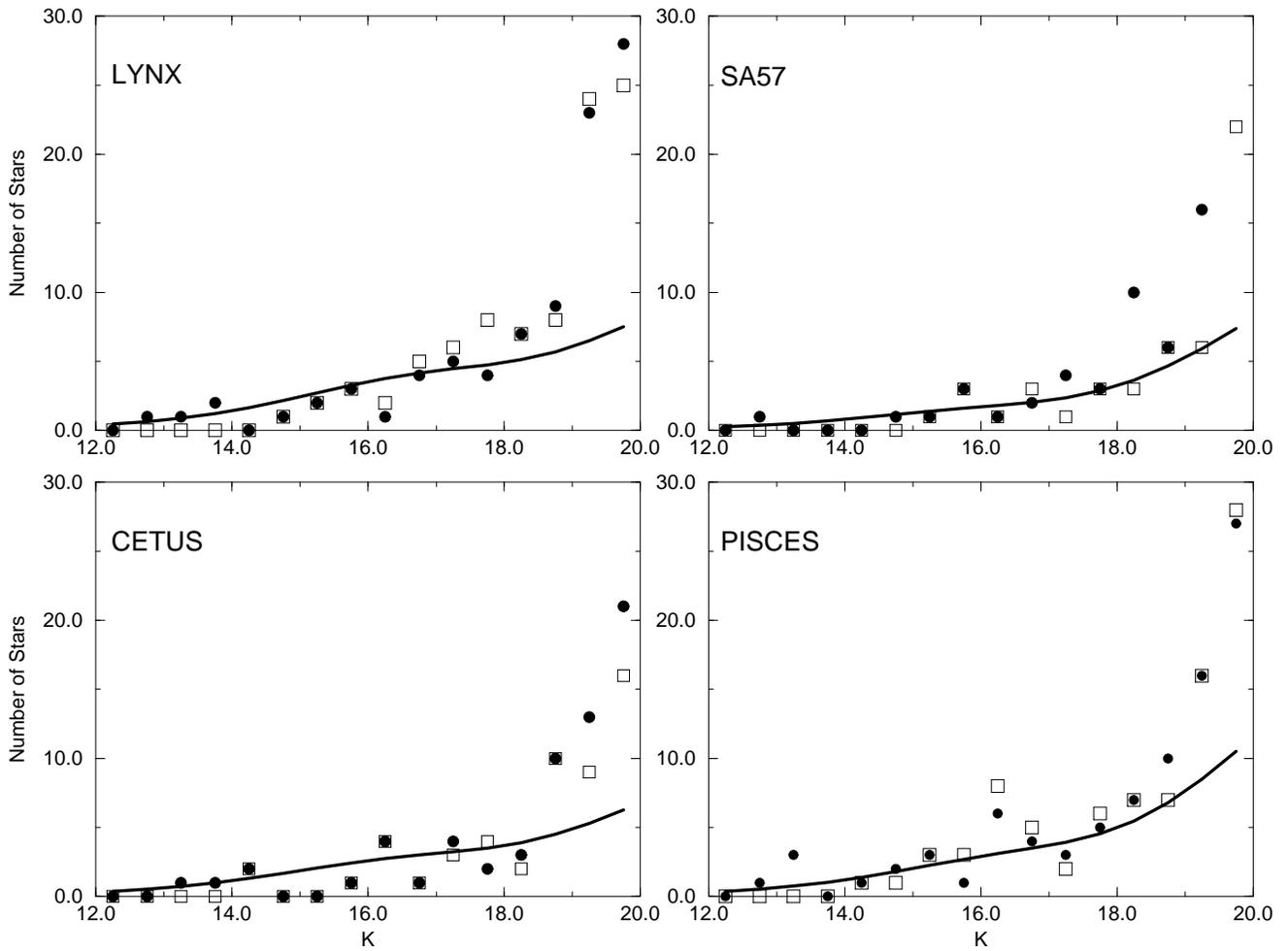}
\epsscale{1.0}
\end{figure}

As described in SED98, we used {\it HST} images to determine
morphologies for objects detected in our $K$--band images of a subset
of the clusters in the whole sample.  For this subsample, we plot
$J-K$ vs. $K$ in Figure~\ref{jkhst}.  As is well known, stars
generally lie in the region $J-K < 1$ (e.g., Leggett 1992) whereas the
$k$--correction for nearly all but the most local galaxies makes their
observed $J-K >1$.  To test this, we compare counts of stars in the
$\sim$100 arcmin$^2$ IR field survey of Elston, Eisenhardt, \&
Stanford (1999; hereinafter EES) selected by the color criterion of
Huang et al.\ (1997) with those selected using our $J-K$ criterion,
and with Galaxy star count models from Bahcall \& Soneira (1980,
1981), as modified for use in the infrared according to Mamon \&
Soneira (1982).  Figure~\ref{jkstars} plots star counts in the four
fields of EES as a function of $K$.  Star counts using our $J-K$
criterion and that of Huang et al.\ (1997) agree with each other and
with the models, down to about $K=18$ (3 magnitudes brighter than the
5 $\sigma$ level in the EES data), where both techniques overpredict
the model star counts.  At $K = 18$, however, the field galaxy counts
outnumber those of the stars in any of the EES fields by $>$10:1,
and thus the details of the stellar contamination correction are
unimportant beyond this magnitude.  We have used our $J-K$
discriminator to eliminate stars more than 3 magnitudes brighter than
the 5 $\sigma$ limit of each cluster's $K$ data, and then switch to
the Bahcall \& Soneira model predictions at fainter magnitudes.  In
the end, the stellar contamination amounts to only a few stars per
half-magnitude bin in each cluster.

\subsection{Galaxy Counts}

We need to correct our cluster galaxy number counts for contamination
by background and foreground galaxies.  This is carried out
statistically, using field galaxy counts from our own observations
(Dickinson et al.\ 1999; EES) and the literature.  Our primary source
is the $K$--band photometry of EES, which has been carried out in
exactly the same manner as for our clusters.

\begin{figure}
\caption{$K$--band galaxy counts from near-IR field surveys (Dickinson et al.\
1999 and Elston et al.\ 1999 -- filled symbols) and from the literature:
the HWS, HMWS and HMDS are from Cowie et al.\ (1993).}
\label{kcounts}
\epsscale{0.8}
\plotone{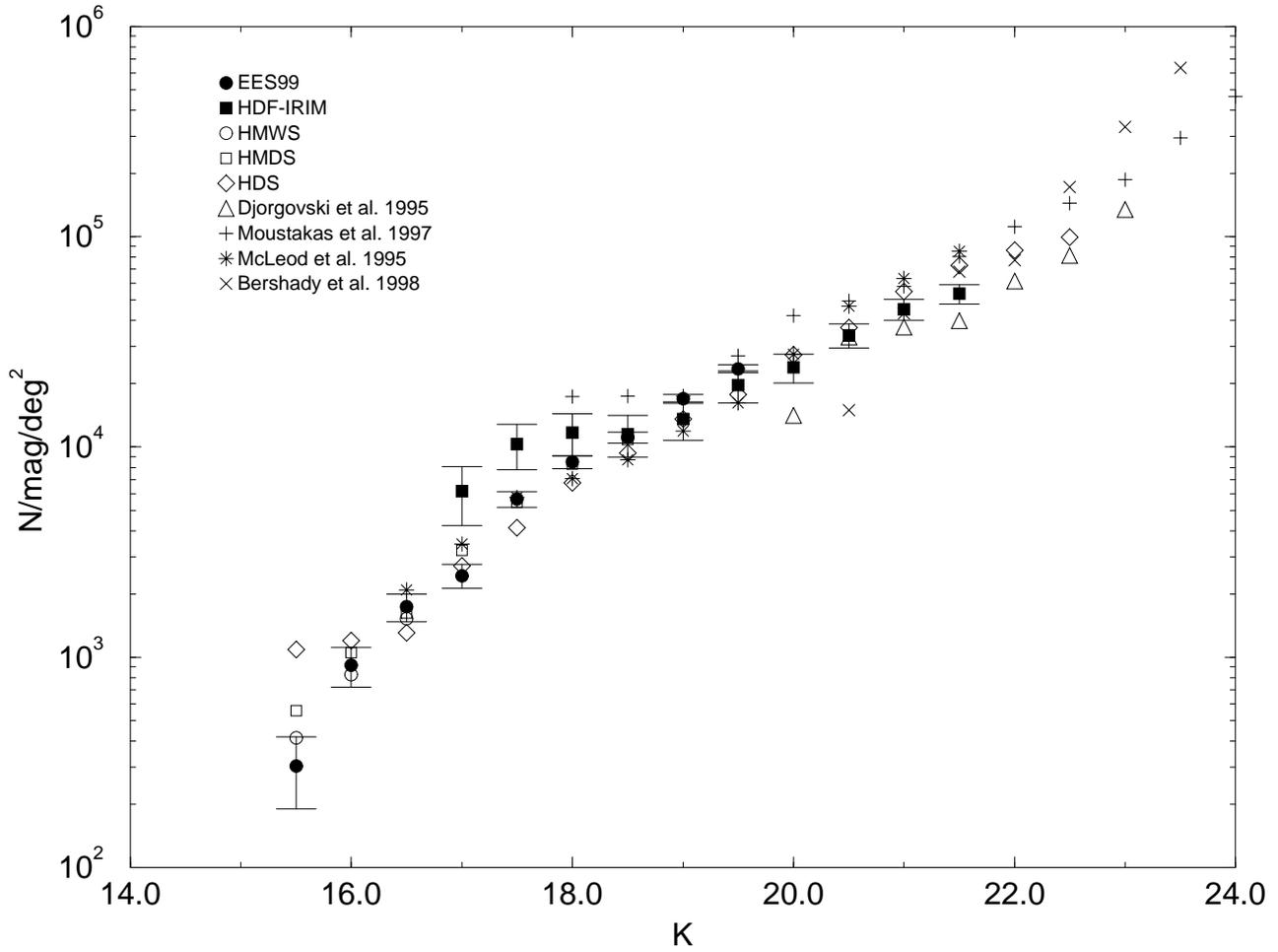}
\epsscale{1.0}
\end{figure}

In Figure~\ref{kcounts} we plot $K$ counts from the EES survey, the
Hubble Deep Field North (HDF--N) using the ground based IR imaging of
Dickinson et al.\ (1999), and a selection of deep $K$ counts from the
literature.  Error bars are calculated following the recipes of Huang
et al.\ (1997) to include an extra contribution due to galaxy
clustering.  The agreement between the various data sets is generally
good, within the variation expected from counting statistics.  There is
an excess in the HDF--N at $17 < K < 19$: comparison of the $J-K$ color
distribution of these galaxies with objects of the same magnitude in
EES shows no difference.  We suggest that this excess is due to a
random enhancement in the numbers of bright galaxies in the small HDF--N
volume.

Next, we obtain linear fits to the $K < 16$ and $K > 18$ counts.  For
$K < 16$ we derive a slope of 0.677, in excellent agreement with the
0.689 of Huang et al.\ (1997).  For $K > 18$ we find a slope of 0.302,
steeper than the 0.261 of Huang et al.\ but in satisfactory agreement
with other $K$--band counts (e.g., Moustakas et al.\ 1997).  Finally,
we correct for field galaxies using the $K < 16$ relation for objects
at $K < 17$, and the $K > 18$ relation for fainter objects, normalized
to the area observed in each cluster.  Errors for these galaxy counts,
used for background subtraction, are calculated according to the
prescription of Huang et al.  After subtracting stars and field
galaxies, the remaining numbers of objects in each cluster, which are
assumed to be member galaxies, are given in Table~\ref{samp} in the
column $N_{memb}$.

\subsection{Luminosity Functions}

We divide our sample of clusters into 10 redshift bins, with central
$z$ between 0.15 and 0.9 as listed in Table 2.  Each bin includes
typically four clusters, but this number varies from bin to bin.  For
example, there are only two clusters in each of the $z=0.15$ and 0.25
bins and so the errors on the fitted $K^*$ are relatively large. For
each cluster we determine the appropriate magnitude intervals based on
the difference between the cluster redshift and the midpoint redshift
of each bin, including the small $k$--correction, so that the intervals
will align at the midpoint redshift.  For example, the $K=16.25 -
16.75$ interval for the midpoint $z=0.397$ bin corresponds to $K=16.38
- 16.89$ for GHO 0303+1706 at $z=0.418$.  We then count the 
objects in the appropriate 0.5 magnitude intervals, removing stars from
the brighter intervals using $J-K$ colors.  Finally faint stars and
field galaxies are removed by interpolating model counts to these
intervals using the methods specified in \S 3.1 and \S 3.2.  Errors in
the raw cluster counts and in the star count models are assumed to be
Poissonian, whereas for contaminating galaxies we add an extra
contribution due to clustering, as in \S 3.2.  We assume errors
propagate in quadrature.

For each redshift bin, we sum number counts for all clusters in order
to reduce shot noise and average over uncertainties in background
subtraction.  We fit the composite LF in each bin with a Schechter
function, fixing the faint end slope to have $\alpha=-0.9$.  This is
the value measured in the infrared LF of the Coma cluster for $K <
K^*+3$ (De Propris et al.\ 1998) and also for the field (Gardner et
al.\ 1997).  We choose not to fit $\alpha$ because our photometry only
reaches to $\sim$2 magnitudes below $L^*$.  The optical LFs for bright
cluster galaxies also have $\alpha \sim -1$, although there is
considerable variation in the slope fainter than four magnitudes below
$L^*$.  We derive values for $K^*$ and the associated errors using the
maximum likelihood technique of Sandage, Tammann, \& Yahil (1979).
Luminosity functions for each of our redshift bins are shown in Figure
5.  The derived values and their $1\sigma$ uncertainties are given
in Table~\ref{lffits}.

Finally, we have divided the sample in two ways in an attempt to
ascertain the effects of cluster mass on $K^\ast (z)$.  First, we
separated by X-ray luminosity (Table~\ref{samp}) at $L_X \sim 4 \times
10^{44}$ ergs$^{-1}$ s$^{-1}$ (0.3--3.5~keV).  This is close to the
characteristic $L_X^*$ of the X-ray luminosity function measured at
similar redshifts (Rosati et al.\ 1998).  X-ray data were
available for 27 clusters from our sample; the remaining 11 were
excluded from this analysis.  X-ray luminosity correlates with total
cluster mass, which may be the main parameter of interest for testing
theoretical models of galaxy formation in clusters.  Schechter
functions were fit separately to the low-- and high--$L_X$ subsamples,
again binned by redshift, with $\alpha$ fixed at $-0.9$.  The fitted
$K^*$ are given in Table~\ref{xraylffits} and the LFs are shown in
Figure~\ref{xraylfbins}.

We also divided the clusters on the basis of ``member'' surface
density as a measure of the cluster richness.  The statistical
$N_{memb}$ were normalized by the field size of the $K$-band images to
calculate the near-IR cluster galaxy density.  For this purpose,
clusters were excluded for which our $K$-band images were relatively
shallow compared to the other data at similar redshift.  The excluded
clusters are MS 1253.9+0456, Cl 2244-02, Abell 370, 3C~313, Vidal 14,
and 3C~34.  The remaining 32 clusters were divided at a ``member''
surface density of 85 Mpc$^{-2}$ into rich and poor groupings.  Again,
Schechter functions were fit separately to the rich and poor
subsamples binned by redshift, with $\alpha$ fixed at $-0.9$.  The
fitted $K^*$ are given in Table~\ref{rplffits} and the LFs are shown
in Figure~\ref{rplfbins}.

\section{Discussion}
The behavior of the characteristic magnitude $K^*$ with redshift is
shown in Figure~\ref{ksvsz}, along with spectral synthesis models for
no--evolution (i.e. pure $k$--correction) and passive evolution
constructed using GISSEL (Bruzual \& Charlot 1993, 1997; hereinafter BC).
All galaxy models used in this paper form stars in a 0.1~Gyr burst
with a Salpeter IMF and with solar metallicity.  The models plotted in
Figure~\ref{ksvsz} were normalized to $K^*=10.9 \pm 0.2$ at the Coma
cluster (De Propris et al.\ 1998), although this normalization was
left free for the statistical tests described below.  The
no--evolution (NE) predictions use model spectra with ages of 10~Gyr
for the $\Omega_M = 1$, $\Lambda = 0$ cosmology, 11~Gyr for $\Omega_M
= 0.3$, $\Lambda=0$, and 12~Gyr for $\Omega_M = 0.3$, $\Lambda = 0.7$.
However, in the near infrared the spectra of galaxies vary so little
that virtually any reasonable model spectrum yields very similar
$k$--corrections.

The observed $K^*$ vs.\ $z$ relation is inconsistent with an
unevolving luminosity function, especially for the low--$\Omega_M$
cosmologies.  The measured $K^*$ values become systematically brighter
at higher redshifts relative to the NE models.  We have tested our
data against the three NE models, including the Coma $K^*$ data point
and allowing for free normalization of the absolute magnitude scaling.
A simple least--squares fit gives the following values for the reduced
$\chi^2$ (for 10 degrees of freedom): 3.0 for $\Omega_M = 1$,
$\Lambda=0$; 4.0 for $\Omega_M = 0.3$, $\Lambda = 0$; and 6.2 for
$\Omega_M = 0.3$, $\Lambda = 0.7$.  The formal probabilities of these
models being consistent with the data are $8.8\times 10^{-4}$,
$1.6\times 10^{-5}$, and $1.5\times 10^{-9}$, respectively.  The
anticorrelation of absolute magnitude with redshift (i.e.\ luminosity
evolution) was examined by applying the Spearman rank--order test to
the difference between the data points and the NE models.  This gives
correlation coefficients of $-0.83$, $-0.86$ and $-0.90$ for the above
three cosmologies.  The probabilities that these anticorrelations
would arise by chance from an unevolving data set are $8.4 \times
10^{-4}$, $4.0 \times 10^{-4}$, and $8.0 \times 10^{-5}$,
respectively.

The passively evolving models were calculated for the $\Omega_M =
0.3$, $\Lambda = 0$ and $\Omega_M = 0.3$, $\Lambda = 0.7$ cosmologies,
assuming $H_0 = 65$~km~s$^{-1}$~Mpc$^{-1}$ and with formation
redshifts $z_f = 2$ and 3.  These provide a better fit to the data
than do the NE models.  We do not attempt to discriminate
statistically between the different evolving models because the
cosmological parameters, formation redshifts, and star formation
histories can be traded off against one another to yield comparably
good fits for different scenarios.  For passively evolving models in
the $\Lambda = 0.7$ cosmology, $z_f = 2$ provides a slightly better
fit than does $z_f = 3$, while for $\Lambda = 0$ the $z_f = 3$ model
is favored.  Extending this analysis to clusters at $z > 1$ should
help to distinguish between the different models.

Barger et al.\ (1998) presented $K$--band LFs for a sample of 10
clusters at $0.31 < z < 0.56$.  Assuming a fixed $\alpha = -1.0$, $H_0
= 50$ km$^{-1}$ s$^{-1}$ Mpc$^{-1}$, and $q_0 = 0.5$, they found
$M^\ast(K_{rest})$ of $-25.41$, $-25.51$, and $-25.29$ for their
redshift bins at $z=0.31, 0.40$, and $0.56$, respectively.  They
claimed no significant luminosity evolution from a comparison of these
values with the $M^\ast(K_{rest}) = -25.1$ determined by Mobasher et
al.\ 1993 for a local {\it field} sample.  However, the expected
departure from no--evolution at $z = 0.56$, the highest redshift bin
in Barger et al., is $\sim$0.5 mag in the observed $K$--band, which is
only about twice the errors in their derived $M^\ast(K_{rest})$.  If
the Barger et al.\ values are compared with $M^\ast(K_{rest}) = -24.8
- 5 \log h_{50}$ which we have measured for the Coma cluster (De
Propris et al.\ 1998), they are consistent with the amount of
passive evolution that we observe.  Another recent study of the
$K$--band LF in distant clusters (Trentham \& Mobasher 1998) appears
to find no evolution, though this is difficult to discern because no
Schechter function fits were made, and their data covered small areas
in only 5 clusters.  By fitting infrared LFs using a uniform data set
that spans a very wide range of redshifts, we have measured evolution
within a single cluster sample without requiring comparisons to
measurements from other published samples which were observed and
analyzed in a different way.

Our key result is the consistency between the observed evolution of
the infrared {\it luminosities} of cluster galaxies and their {\it
color} evolution as manifested by the slope, scatter and intercept of
the color--magnitude relation (Arag\'on-Salamanca et al.\ 1993; SED95;
SED98; Ellis et al.\ 1997).  Both are fully consistent with simple,
passive spectrophotometric evolution of a galaxy population which has
been largely stable throughout the redshift range $0 < z < 1$.
Kauffmann \& Charlot (1998a) have shown that the apparently passive
evolution of the color--magnitude relation can be accommodated within
a hierarchical model, even if the galaxies themselves still grow by
mergers until late times.  Our new measurements show that galaxies in
high redshift clusters also follow the same intrinsic luminosity
distribution as do those today, once passive evolution is taken into
account, strongly suggesting that their bulk stellar masses have not
increased substantially since $z = 1$.  It is interesting to note that
Arag\'on--Salamanca et al.\ (1998) find that the {\it brightest}
cluster galaxies (BCGs) exhibit a rather different behavior, with
infrared luminosities which are {\it fainter} at high redshift once
the expected evolution of their stellar populations is taken into
account.  They interpret this as evidence that BCGs have grown with
time by accretion, a behavior which we would suggest is not shared by
the bulk of the cluster population.
 
Previous photometric studies of early--type cluster galaxies at other
wavelengths have reached conclusions that are qualitatively similar to
our own.  The rest frame $B$--band surface brightnesses of cluster
ellipticals at $z \leq 1.2$ have been shown to be consistent with
high-$z$ formation and passive evolution (Dickinson 1995, 1997; Pahre,
Djorgovski \& De~Carvalho 1996; Schade, Barrientos \& Lopez-Cruz 1997;
Barger et al.\ 1998), albeit with large uncertainties.  Moreover,
recent studies exploiting the full power of the Fundamental Plane now
probe nearly the same redshift range as do our LFs (e.g.\ Kelson et
al.\ 1997; van Dokkum et al.\ 1998).  These have shown that the
rest--frame $B$--band mass--to--light ratio ($M/L$) of early--type
cluster galaxies evolved in a manner consistent with passive
luminosity evolution.  Early results derived from a small sample at
$z=0.83$ indicate a limit on the formation redshift for cluster
ellipticals of $z_f > 1.7$ in an $\Omega_M = 0.3$, $\Omega_{\Lambda} =
0.7$ universe (van Dokkum et al.\ 1998).  If the evolution of both
$M/L$ and of the luminosity function can be consistently explained by
passive stellar evolution, then it would follow that the {\it mass
function} of cluster galaxies is approximately invariant over the
range of redshifts and luminosities which have been studied to date.

The luminosity evolution that we observe in our cluster sample is
strikingly different from that which has been reported for {\it field}
galaxy samples, particularly those selected in the near--infrared or
limited to early--type galaxies only.  Cowie et al.\ (1996) present
rest--frame $K$--band luminosity functions derived from a field galaxy
sample spanning $0 < z < 1.6$, and find that $L^*_K$ remains constant
or declines at higher redshifts.  Similarly, Kauffmann, Charlot, \&
White (1996) cite evidence for a strong negative luminosity evolution
of early--type field galaxies selected by color from the
Canada--France Redshift Survey (Lilly et al.\ 1995a).  Kauffmann \&
Charlot (1998b) have highlighted the deficit of high redshift galaxies
at $K < 19$ in the infrared--selected surveys of Songaila et al.\
(1994) and Cowie et al.\ (1996).

This apparent deficit of bright, high redshift field galaxies has been
cited as evidence supporting hierarchical models which would assemble
the most massive galaxies at relatively late epochs.  This
interpretation is not necessarily inconsistent with our measurements,
since field and cluster galaxies may follow different evolutionary
histories.  Indeed, in the semi--analytic models of Kauffmann \&
Charlot (1998a), cluster ellipticals form earlier than field
ellipticals, and might therefore be expected to show different
luminosity evolution.  At some level, if the assembly of galaxies is
pushed back to higher redshifts in denser environments, then it is
inevitable that their subsequent behavior will more closely resemble
passive evolution at late times, and the distinction between
``passive'' and ``hierarchical'' models will blur.

There is conflicting evidence on the question of the dependence of
elliptical galaxy evolution on environment.  Schade et al.\ (1996)
find no difference in the $M_B$-$\log r_e$ relations of distant field
and cluster ellipticals.  Bernardi et al.\ (1998) find nearly
identical Mg$_2$-$\sigma$ relations for field and cluster ellipticals
from large, nearby galaxy samples, and use this to limit mean age
differences to be $< 1$ Gyr, arguing that the bulk of stars in
galactic spheroids was formed at high redshift, independent of
environmental density.  This is also the conclusion reached by De
Propris et al.\ (1998) on the basis of the very similar infrared
luminosity functions of field and cluster galaxies.  On the other
hand, Worthey and collaborators (e.g., Worthey 1997 and references
therein) have suggested that field E/S0 galaxies span a large range of
ages, on the basis of Balmer line strengths in their integrated
spectra.  This is not necessarily inconsistent with our findings,
since relatively small bursts suffice to account for the observations.
Conversely, far-ultraviolet components may significantly affect the
use of narrow band spectral indices as age indicators (Davies, Sadler,
\& Peletier 1993).

Kauffmann \& Charlot (1998a) suggest that studies of galaxies in rich
clusters are strongly biased, as they select massive objects whose
members are likely to have formed at high redshift in any scenario for
structure formation.  They argue that the evidence for passive
evolution comes from selecting samples which are most likely to be
passively evolving.  Our sample includes clusters which span two
orders of magnitude in X-ray luminosity, with a wide range of optical
richness, and presumably also of mass.  As described in \S3.3, we have
divided our cluster data into subsamples with high and low X-ray
luminosity, and with rich and poor ``member'' density, in order to
test whether the luminosity evolution depends on the mass and/or
optical richness of the cluster.  The behavior of $K^*$ vs.\ $z$ for
the two X-ray subsamples (Figure~\ref{ksvszlx}) and for the rich vs
poor groupings (Figure~\ref{ksvszrp}) appear to be similar in each
case.  The only exception is in the highest redshift bin in the rich
vs poor comparisons, where $K^*$ for the rich clusters is brighter
than that of the poor clusters at about the 2$\sigma$ level.  Overall,
these two methods both indicate that the luminosity evolution of
massive galaxies is similar across the range of cluster environments
spanned by our sample, and that biases of the type suggested by
Kauffmann \& Charlot (1998a) are likely to be weak, at least for
environments richer than groups.

The LFs presented here sum the entire population of galaxies within a
radius of $\sim$0.5 to 1 Mpc from the cluster centers.  For $K < K^* +
2$, this is likely to be dominated by E/S0 galaxies and early type
spirals, although the morphology--density and morphology--radius
relationships appear to be weaker in clusters at high redshift
(Dressler et al.\ 1997).  It is unclear how the changing morphological
mix of galaxies in distant clusters might affect the $K$--band
luminosity functions, but the presence of an increasingly numerous
population of blue, star--forming cluster galaxies at higher redshifts
does not fit well within any simple model of pure, passive evolution.
Since the infrared LF primarily measures the distribution of stellar
mass among galaxies, the apparently passive luminosity evolution and
``active'' Butcher--Oemler population might be reconciled if the
starbursts involve only a small fraction of the total mass in each
galaxy.  Indeed, population synthesis models show that even relatively
small amounts of star formation, in terms of the total stellar mass
fraction, can account for the colors of most blue galaxies in distant
clusters (Barger et al.\ 1996).

\section{Conclusions}

We have presented $K$--band luminosity functions for a heterogeneous
sample of 38 galaxy clusters spread over $0.1 < z < 1$.  Schechter
function fits to the field--corrected galaxy counts yield $K^*$ values
brighter than would be expected for a non--evolving population of
early--type galaxies at $z > 0.4$ in low $\Omega$ cosmologies, and
consistent with pure, passive luminosity evolution.  This result
supports and extends our previous results based on analysis of the
color evolution of early--type cluster galaxies over the same redshift
range.  The positive luminosity evolution for cluster galaxies appears
to differ from the flat or negative evolution which has been reported
for infrared-- and color--selected {\it field} galaxy samples out to
$z \approx 1$ (Cowie et al.\ 1996; Kauffmann, Charlot \& White 1996;
Kauffmann \& Charlot 1998b), perhaps suggesting a different
evolutionary history for massive galaxies in different environments.
However, we do not observe different evolution when we divide our
cluster sample by X-ray luminosity or by richness.

These observations point to the importance of obtaining a census of 
galaxy properties over a wide range of environments for understanding
the mechanisms which drive galaxy evolution.  Hierarchical models in
which collisionless dark matter dominates galaxy dynamics and
evolution generally postulate that the formation of massive galaxies,
such as cluster ellipticals, takes place by accretion and mergers that
continue to low redshift.  Our results indicate that the stellar
populations of massive galaxies ($K < K^* + 2$) in cluster cores form
at relatively high redshifts, $z \gtrsim 2$, and suggest that the
assembly of those galaxies is largely complete by $z \sim 1$.

\acknowledgments

The authors would like to thank NOAO for a generous allocation of
observing time to this project, and the staffs at Kitt Peak and Cerro
Tololo for their help with the observing.  This research has made use
of the NASA/IPAC Extragalactic Database (NED) which is operated by the
Jet Propulsion Laboratory, California Institute of Technology, under
contract with the National Aeronautics and Space Administration.
Support for this work was provided by NASA through grant number
AR--5790.02--94A from the Space Telescope Science Institute, which is
operated by the Association of Universities for Research in Astronomy,
Inc., under NASA contract NAS5-26555.  Funding for OSIRIS was provided
by grants from The Ohio State University and National Science
Foundation grants AST--9016112 and AST--9218449.  Portions of the
research described here were carried out by the Jet Propulsion
Laboratory, California Institute of Technology, under a contract with
NASA.  Work performed at the Lawrence Livermore National Laboratory is
supported by the DOE under contract W7405-ENG-48.  Work at the
University of New South Wales is supported by a grant from the
Australian Research Council.  We would like to thank an anonymous
referee for a number of helpful suggestions.

\clearpage

\begin{deluxetable}{lccccccccc}
\small
\tablecaption{Cluster Sample}
\tablewidth{7.3in}
\tablehead{
\colhead{Name} & \colhead{R.A.} & \colhead{Dec.} & \colhead{Data\tablenotemark{a}} & \colhead{Field}
& \colhead{$z$} & \colhead{$K_{lim}$} & \colhead{N$_{samp}$} 
& \colhead{N$^b_{memb}$} & \colhead{L$_x$ } 
\\
\colhead{} & \colhead{J2000} & \colhead{J2000} & \colhead{} & \colhead{arcmin$^2$}
& \colhead{} & \colhead{mag} & \colhead{} 
& \colhead{} & \colhead{$10^{45}$ ergs/s}
}
\startdata
Abell 1146	& 11:01:20.6 & $-$22:43:08 &4& 59.3 & 0.142 & 16.5 & 187 & 129 & 0.33 \nl
Abell 3305	& 05:01:52.9 & $-$39:12:45 &4& 52.9 & 0.157 & 16.5 & 100 &  46 & \nodata \nl
MS 0906.5+1110	& 09:09:16.7 & $+$10:58:38 &4& 57.5 & 0.180 & 16.9 & 207 & 146 & 0.57 \nl
Abell 1689	& 13:11:34.2 & $-$01:21:56 &4& 54.1 & 0.182 & 17.4 & 259 & 176 & 1.30 \nl
Abell 1942	& 14:38:37.0 & $+$03:40:05 &4& 15.9 & 0.224 & 17.5 & 121 &  96 & 0.16 \nl
MS 1253.9+0456	& 12:56:28.8 & $+$04:40:02 &4& 16.1 & 0.230 & 17.2 & 117 &  94 & 0.52 \nl
Abell 1525	& 12:22:03.8 & $-$01:08:38 &4& 16.0 & 0.259 & 17.5 & 85  &  55 & \nodata \nl
MS 1008.1-1224	& 10:10:34.1 & $-$12:39:48 &4& 17.2 & 0.301 & 17.9 & 207 & 154 & 0.43 \nl
MS 1147.3+1103	& 11:49:55.7 & $+$10:46:37 &4& 14.9 & 0.303 & 17.9 & 115 &  79 & 0.38 \nl
AC 118		& 00:14:19.3 & $-$30:23:18 &3& 24.4 & 0.308 & 18.5 & 292 & 200 & 1.60 \nl
AC 103		& 20:57:07.5 & $-$64:38:53 &3& 23.8 & 0.311 & 18.2 & 277 & 203 & \nodata \nl
AC 114		& 22:58:52.0 & $-$34:46:54 &3& 21.8 & 0.312 & 18.3 & 276 & 161 & 0.50 \nl
MS 2137.3-234 	& 21:40:14.5 & $-$23:39:41 &3& 24.3 & 0.313 & 18.1 & 184 & 135 & 1.50 \nl
Abell S0506	& 05:01:04.0 & $-$24:24:42 &4& 18.4 & 0.316 & 18.0 & 150 &  92 & 0.13 \nl
MS 1358.1+6245	& 13:59:54.3 & $+$62:30:36 &1& 21.5 & 0.328 & 18.2 & 179 & 116 & 1.06 \nl
Cl 2244-02 	& 22:47:12.9 & $-$02:05:40 &1& 21.5 & 0.330 & 18.8 & 213 & 114 & 0.50 \nl
Abell 370 	& 02:39:53.8 & $-$01:34:24 &5& 22.4 & 0.374 & 18.1 & 190 & 145 & 1.10 \nl
Cl 0024+16 	& 00:26:35.4 & $+$17:09:51 &1& 21.4 & 0.391 & 18.8 & 329 & 230 & 0.22 \nl 
Abell 851 	& 09:43:02.6 & $+$46:58:37 &5& 21.5 & 0.405 & 18.7 & 306 & 196 & 0.80 \nl
GHO 0303+1706 	& 03:06:15.9 & $+$17:19:17 &1& 21.3 & 0.418 & 18.8 & 252 & 186 &  0.22 \nl
3C~313		& 15:10:59.6 & $+$07:51:49 &1& 21.3 & 0.461 & 18.5 & 158 &  60 & \nodata \nl
3C~295		& 14:11:19.5 & $+$52:12:21 &1& 21.9 & 0.461 & 18.8 & 222 & 133 & \nodata \nl
\tablebreak								      
F1557.19TC 	& 04:12:51.6 & $-$65:50:17 &3& 24.1 & 0.510 & 19.1 & 231 & 131 & 0.05 \nl
Vidal 14	& 00:49:11.1 & $-$24:40:55 &3& 22.1 & 0.520 & 18.0 & 143 &  90 & \nodata \nl 
GHO 1601+4253 	& 16:03:10.6 & $+$42:45:35 &1& 21.6 & 0.539 & 19.2 & 264 & 165 & 0.14 \nl
MS 0451.6-0306 	& 04:54:10.8 & $-$03:00:57 &2&  6.6 & 0.539 & 19.2 & 153 & 110 & 1.90 \nl
Cl 0016+16 	& 00:18:33.6 & $+$16:25:46 &1& 21.7 & 0.545 & 19.1 & 338 & 241 & 1.60 \nl
J1888.16CL 	& 00:56:54.6 & $-$27:40:31 &3& 21.1 & 0.560 & 19.2 & 253 & 156 & 0.12 \nl
MS 2053.7-0449	& 20:56:22.4 & $-$04:37:43 &3& 27.7 & 0.582 & 19.2 & 443 & 215 & 0.56 \nl
GHO 0317+1521	& 03:20:02.3 & $+$15:31:49 &2&  7.4 & 0.583 & 19.2 & 72  &   7 & \nodata  \nl
3C 220.1	& 09:32:39.6 & $+$79:06:32 &1& 20.2 & 0.620 & 19.5 & 245 & 209 & \nodata \nl
GHO 2201+0258	& 22:04:05.7 & $+$03:12:50 &2&  8.1 & 0.640 & 19.3 & 93  &  31 & \nodata \nl
3C 34 		& 01:10:18.5 & $+$31:47:20 &2&  6.5 & 0.689 & 19.1 & 156 & 109 & \nodata \nl 
GHO 1322+3027 	& 13:24:49.3 & $+$30:11:28 &2&  6.5 & 0.751 & 20.3 & 163 & 112 & 0.09 \nl 
MS 1137.5+6625	& 11:40:23.3 & $+$66:08:41 &2&  7.1 & 0.782 & 20.0 & 135 &  48 & 0.75 \nl
MS 1054.5-032 	& 10:56:59.5 & $-$03:37:28 &2&  6.5 & 0.828 & 20.3 & 201 & 160 & 0.90 \nl
GHO 1603+4313 	& 16:04:18.9 & $+$43:04:36 &2&  6.9 & 0.895 & 20.3 & 144 &  59 & 0.12 \nl
GHO 1603+4329	& 16:04:31.5 & $+$43:21:17 &2&  7.2 & 0.920 & 20.1 & 150 &  59 & \nodata \nl

\enddata
\tablenotetext{a}{1 = KPNO 2.1m and IRIM; 2 = KPNO 4m and IRIM; 3 =
CTIO 1.5m and CIRIM; 4 = CTIO 1.5m and OSIRIS; 5 = KPNO 1.3m and SQIID}
\tablenotetext{b}{Number of cluster ``members'' left, in a statistical
sense, after removing contaminating stars and galaxies following the
methods outlined in \S 3}
\label{samp}
\end{deluxetable}

\begin{deluxetable}{llc}
\tablewidth{3in}
\tablecaption{$K^*$ vs. $z$ for the complete cluster sample}
\tablehead{
\colhead{Redshift} & \colhead{$K^*$} & \colhead{Clusters/bin}
}

\startdata
0.15 & $14.84 \pm 0.49$ & 2 \nl
0.20 & $15.16 \pm 0.07$ & 3 \nl
0.25 & $15.64 \pm 0.38$ & 2 \nl
0.32 & $15.74 \pm 0.08$ & 9 \nl
0.40 & $16.50 \pm 0.11$ & 4 \nl
0.46 & $16.38 \pm 0.08$ & 2 \nl
0.54 & $16.85 \pm 0.18$ & 6 \nl
0.61 & $17.57 \pm 0.41$ & 4 \nl
0.79 & $17.51 \pm 0.26$ & 4 \nl
0.90 & $18.05 \pm 0.25$ & 2 \nl
\enddata
\label{lffits}
\end{deluxetable}

\begin{deluxetable}{llclc}
\tablecaption{$K^*$ vs. $z$ for high and low X--ray luminosity subsamples}
\tablehead{
\colhead{Redshift} & \colhead{$K^*$(high $L_X$)} & \colhead{Clusters/bin} & 
\colhead{$K^*$(low $L_X$)} & \colhead{Clusters/bin}
}

\startdata
0.20 & $15.15 \pm 0.22$ & 3 & $15.08 \pm 0.15$ & 2 \nl
0.32 & $15.64 \pm 0.12$ & 5 & $15.89 \pm 0.27$ & 3 \nl
0.40 & $15.77 \pm 0.28$ & 2 & $16.09 \pm 0.14$ & 2 \nl
0.54 & $16.81 \pm 0.15$ & 3 & $16.88 \pm 0.22$ & 3 \nl
0.83 & $17.64 \pm 0.31$ & 2 & $17.93 \pm 0.10$ & 2 \nl
\enddata
\label{xraylffits}
\end{deluxetable}

\begin{deluxetable}{llclc}
\tablecaption{$K^*$ vs. $z$ for rich and poor subsamples}
\tablehead{
\colhead{Redshift} & \colhead{$K^*$(rich)} & \colhead{Clusters/bin} & 
\colhead{$K^*$(poor)} & \colhead{Clusters/bin}
}

\startdata
0.20 & $15.00 \pm 0.11$ & 3 & $15.25 \pm 0.41$ & 2 \nl
0.32 & $15.77 \pm 0.16$ & 4 & $15.53 \pm 0.11$ & 6 \nl
0.40 & $16.03 \pm 0.16$ & 2 & $16.28 \pm 0.13$ & 2 \nl
0.54 & $16.75 \pm 0.18$ & 2 & $16.50 \pm 0.22$ & 6 \nl
0.83 & $17.21 \pm 0.28$ & 2 & $18.05 \pm 0.25$ & 3 \nl
\enddata
\label{rplffits}
\end{deluxetable}
\clearpage

\begin{figure}
\caption{Cumulative $K$--band luminosity functions for all redshift
bins.  Solid lines are the Schechter function fits with $\alpha = -0.9$.}
\label{lfbins}
\epsscale{0.8}
\plotone{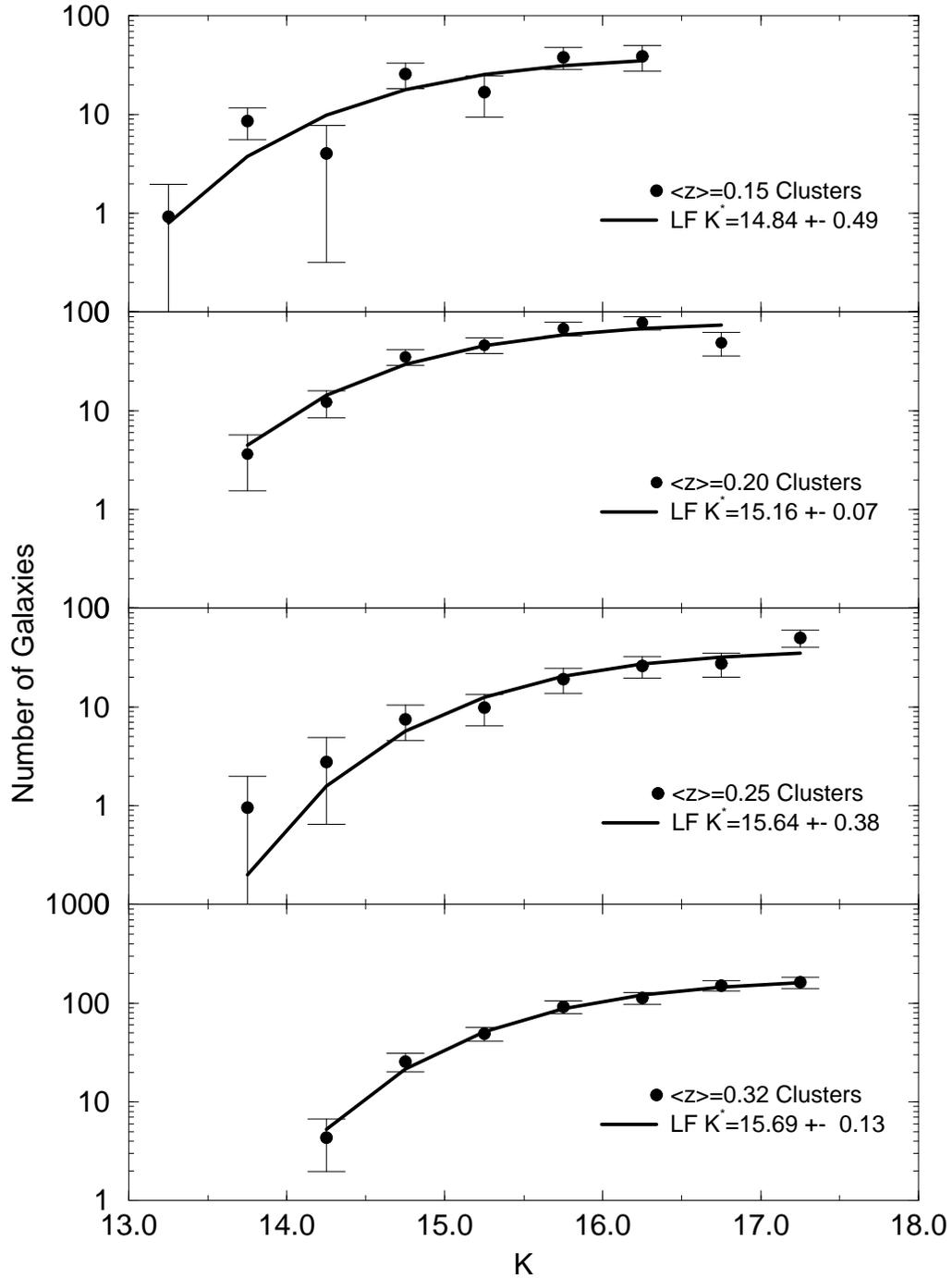}
\end{figure}

\begin{figure}
\figurenum{5}
\caption{(continued)}
\plotone{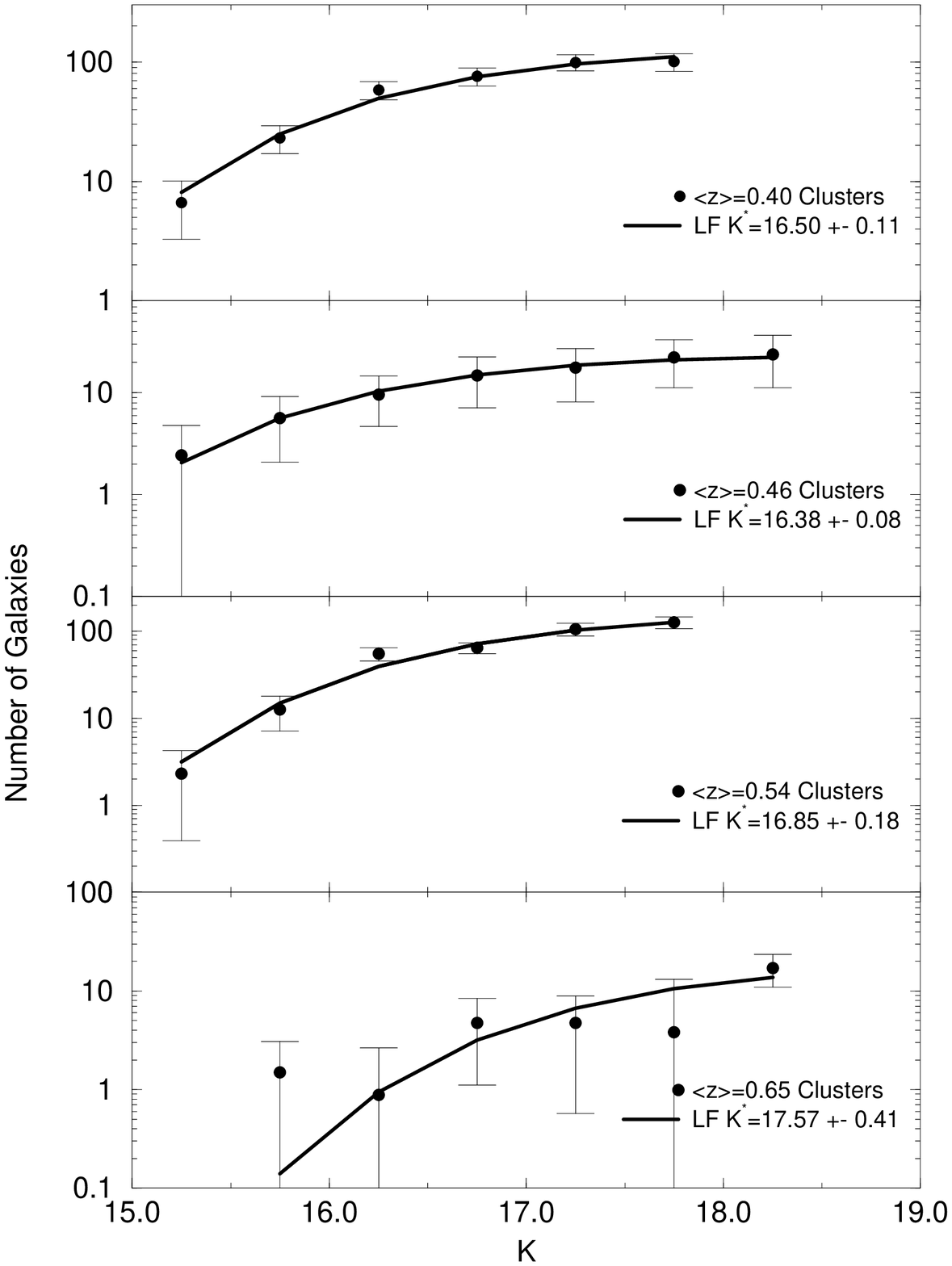}
\end{figure}

\begin{figure}
\figurenum{5}
\caption{(continued)}
\plotone{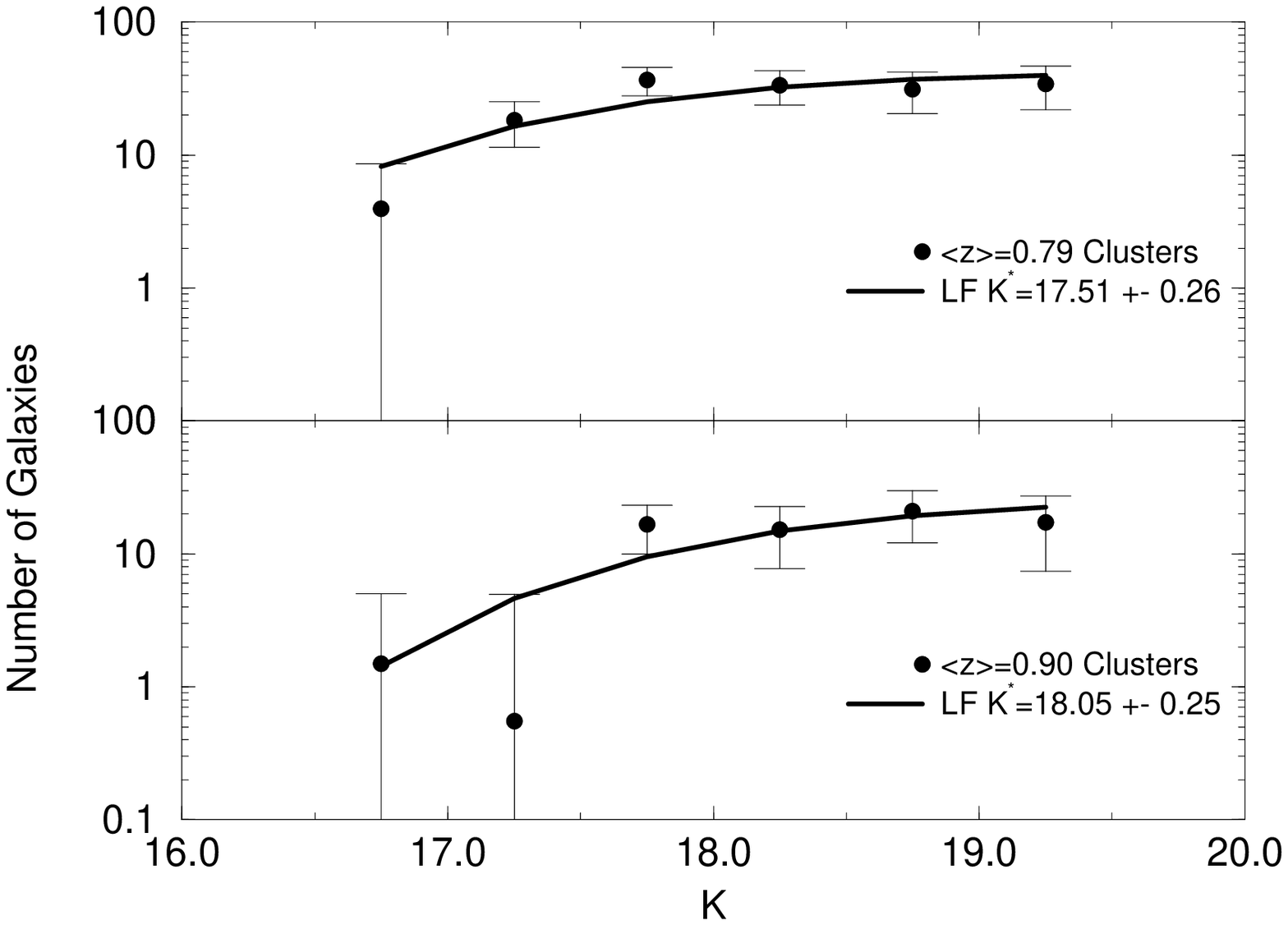}
\epsscale{1.0}
\end{figure}

\begin{figure}
\epsscale{0.8}
\caption{Cumulative luminosity functions in five redshift bins
for high and low X-ray luminosity cluster subsamples,
chosen as described in the text.  The solid and dashed lines 
are the Schechter function fits with $\alpha = -0.9$.}
\label{xraylfbins}
\plotone{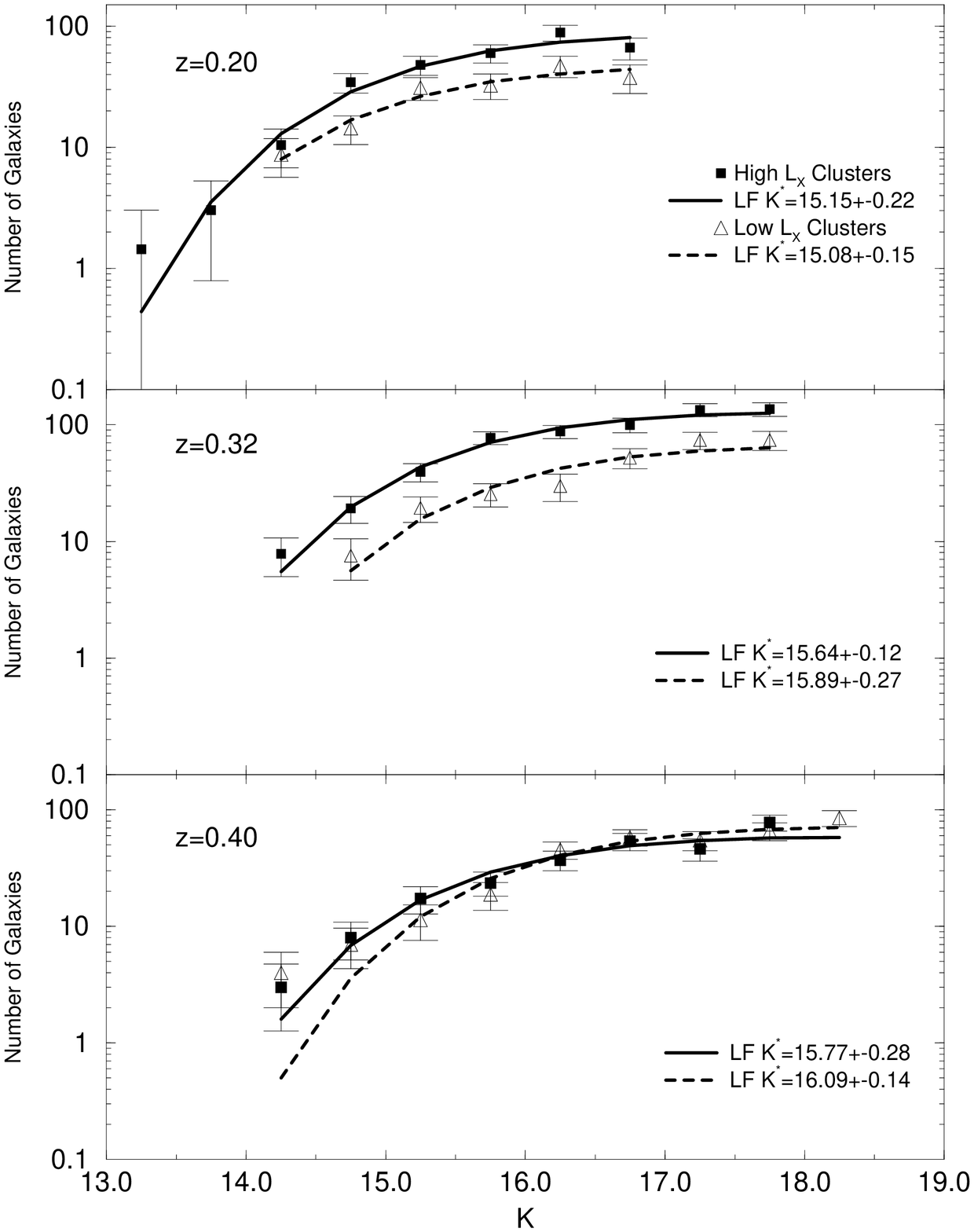}
\end{figure}

\begin{figure}
\figurenum{6}
\caption{(continued)}
\plotone{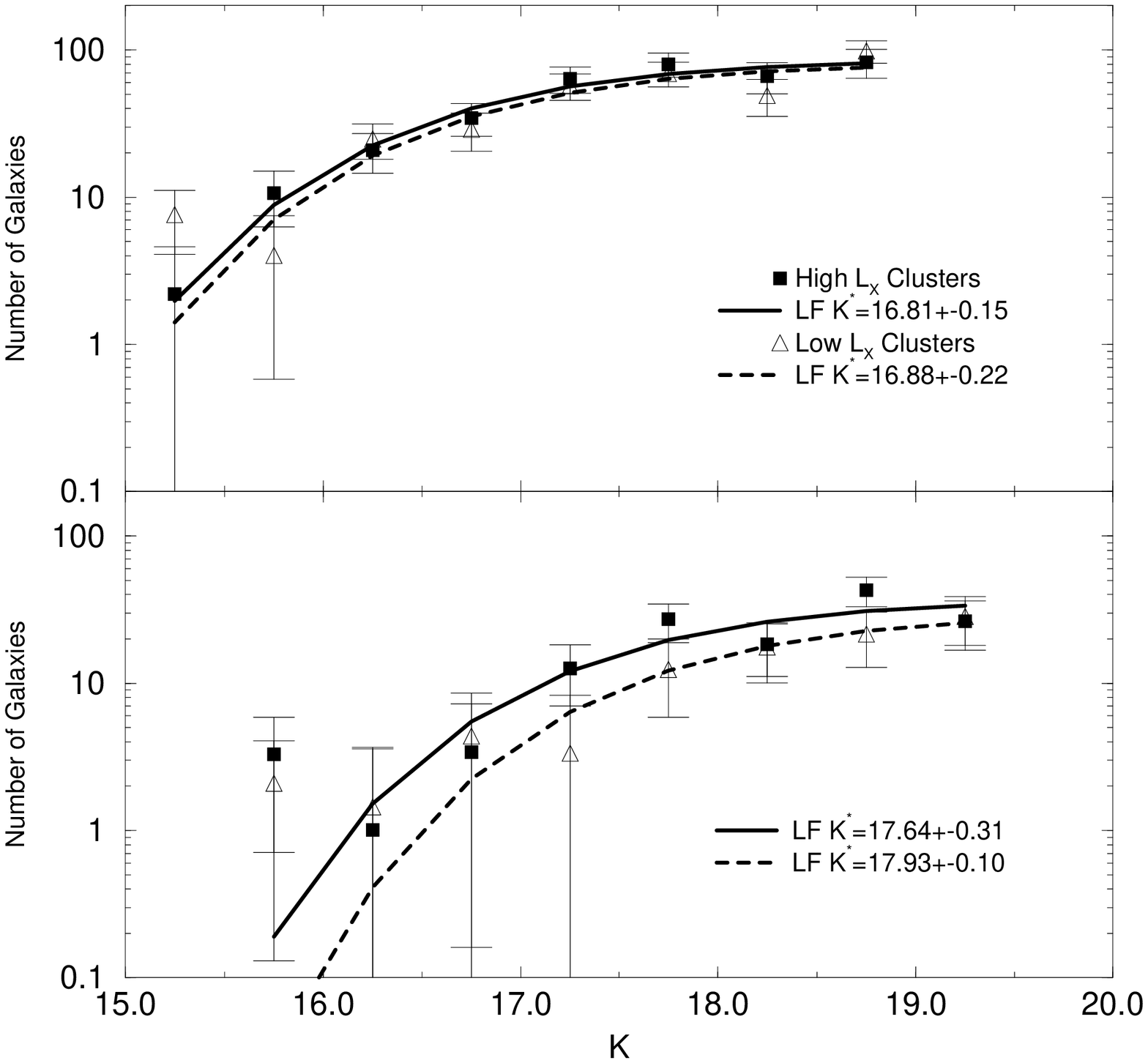}
\epsscale{1.0}
\end{figure}
\clearpage

\begin{figure}
\epsscale{0.8}
\caption{Cumulative luminosity functions in five redshift bins
for the rich and poor cluster subsamples,
chosen as described in the text.  The solid and dashed lines 
are the Schechter function fits with $\alpha = -0.9$.}
\label{rplfbins}
\plotone{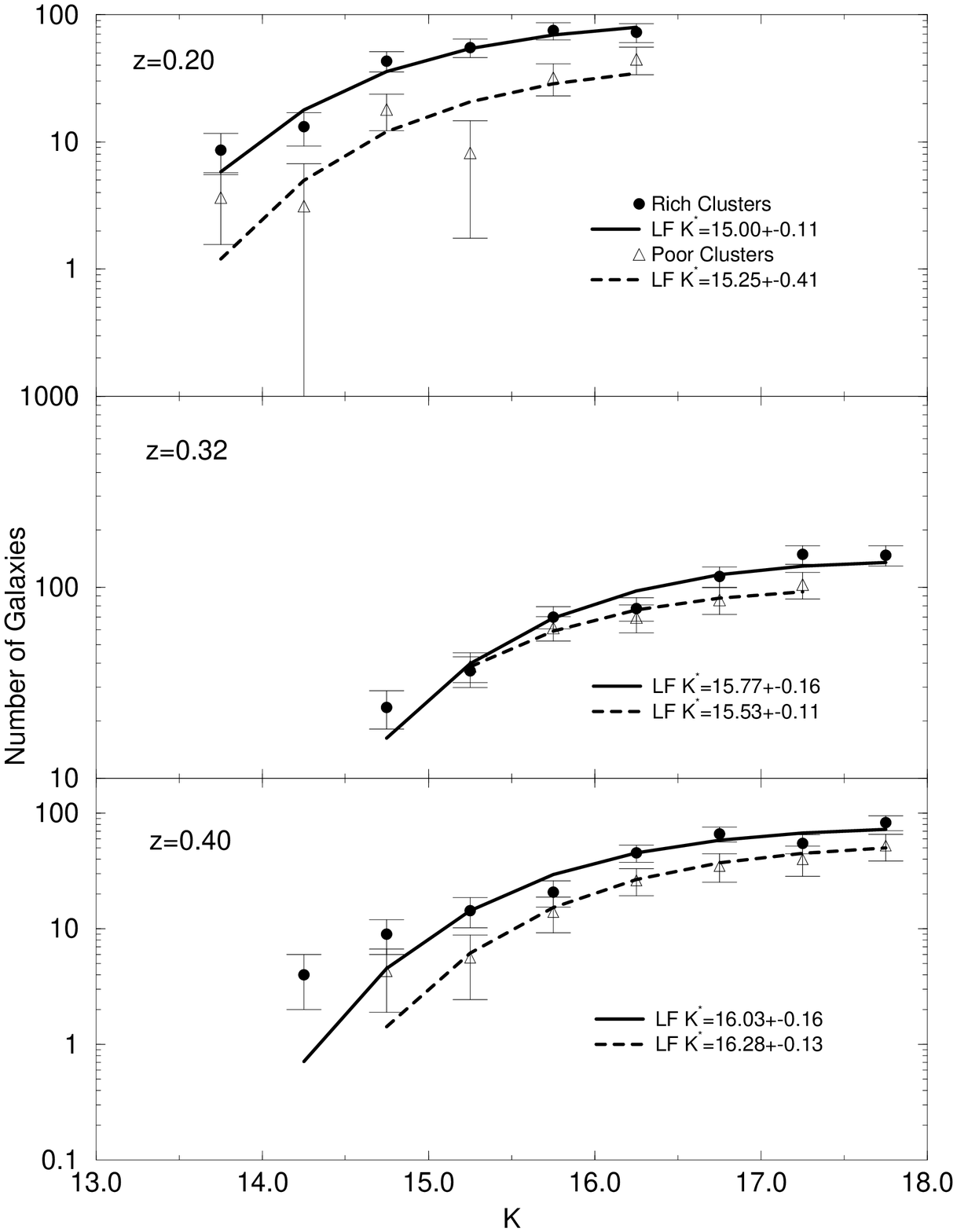}
\epsscale{1.0}
\end{figure}

\begin{figure}
\figurenum{7}
\epsscale{0.8}
\caption{(continued)}
\plotone{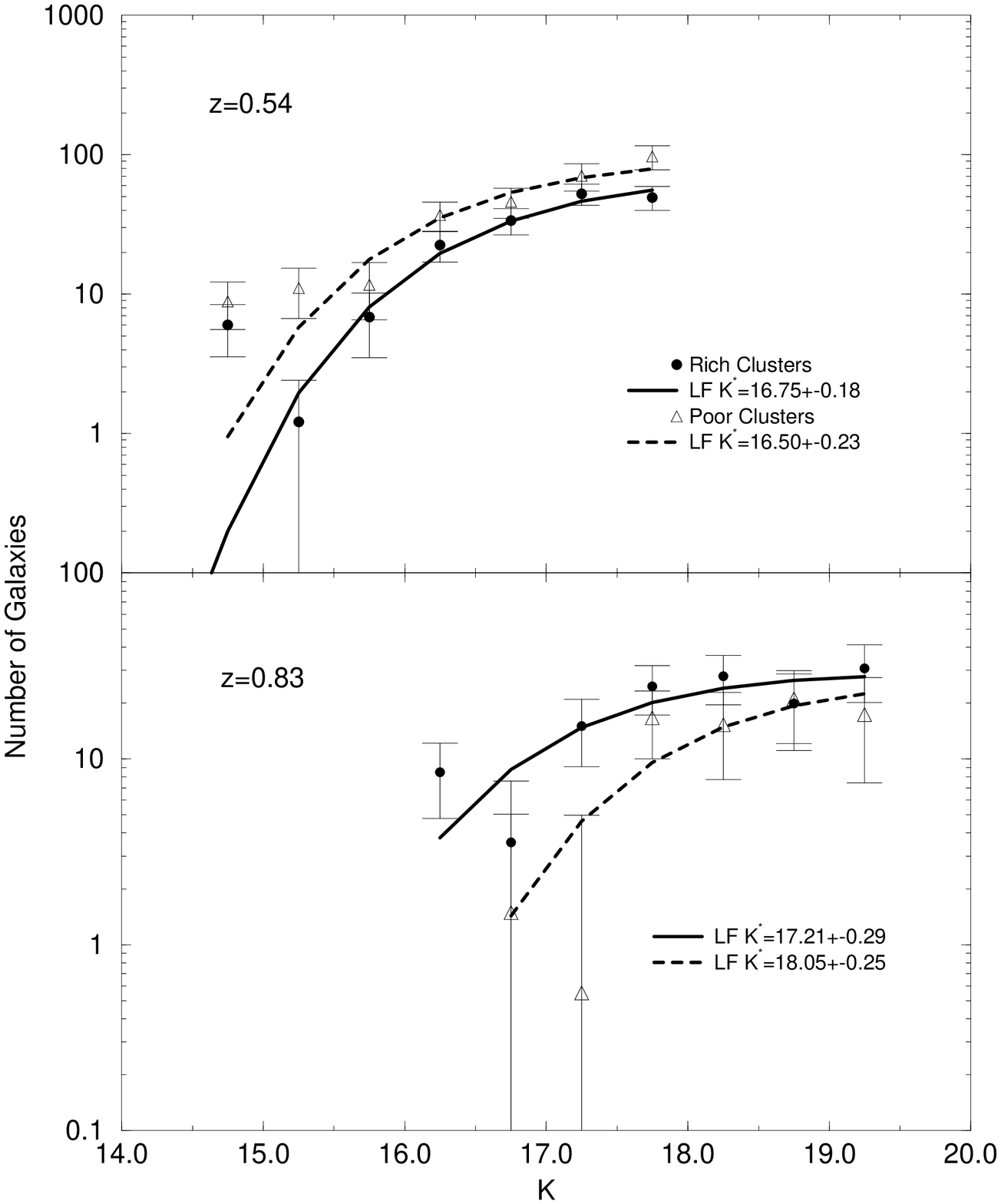}
\epsscale{1.0}
\end{figure}
\clearpage

\begin{figure}
\epsscale{1.1}
\caption{$K^*$ vs. $z$ for the cluster sample binned by
redshift. Lines represent galaxy models determined from the Bruzual
\& Charlot GISSEL, normalized to the Coma cluster which has $K^*=10.9$
(De Propris et al.\ 1998).  These models represent 0.1 Gyr starbursts
with a Salpeter IMF and $Z_\odot$, with $H_0 = 65$ km s$^{-1}$
Mpc$^{-1}$.  For the no-evolution cases, the thick solid line is for
$\Omega_M = 1$ and $\Lambda=0$, the thick dotted line is for $\Omega_M
= 0.3$ and $\Lambda = 0.7$, and the thick dashed line is for $\Omega_M
= 0.3$ and $\Lambda = 0$.  For the passive evolution models: the thin
solid line represents $z_f = 3.0, \Omega_M = 0.3, \Lambda = 0.7$; the
thin dotted line $z_f = 2.0$, $\Omega_M = 0.3, \Lambda = 0.7$; the
thin dashed line represents $z_f = 3.0, \Omega_M = 0.3, \Lambda = 0$;
and the thin dot-dash line $z_f = 2.0$, $\Omega_M = 0.3, \Lambda = 0$.
}
\label{ksvsz} 
\plotone{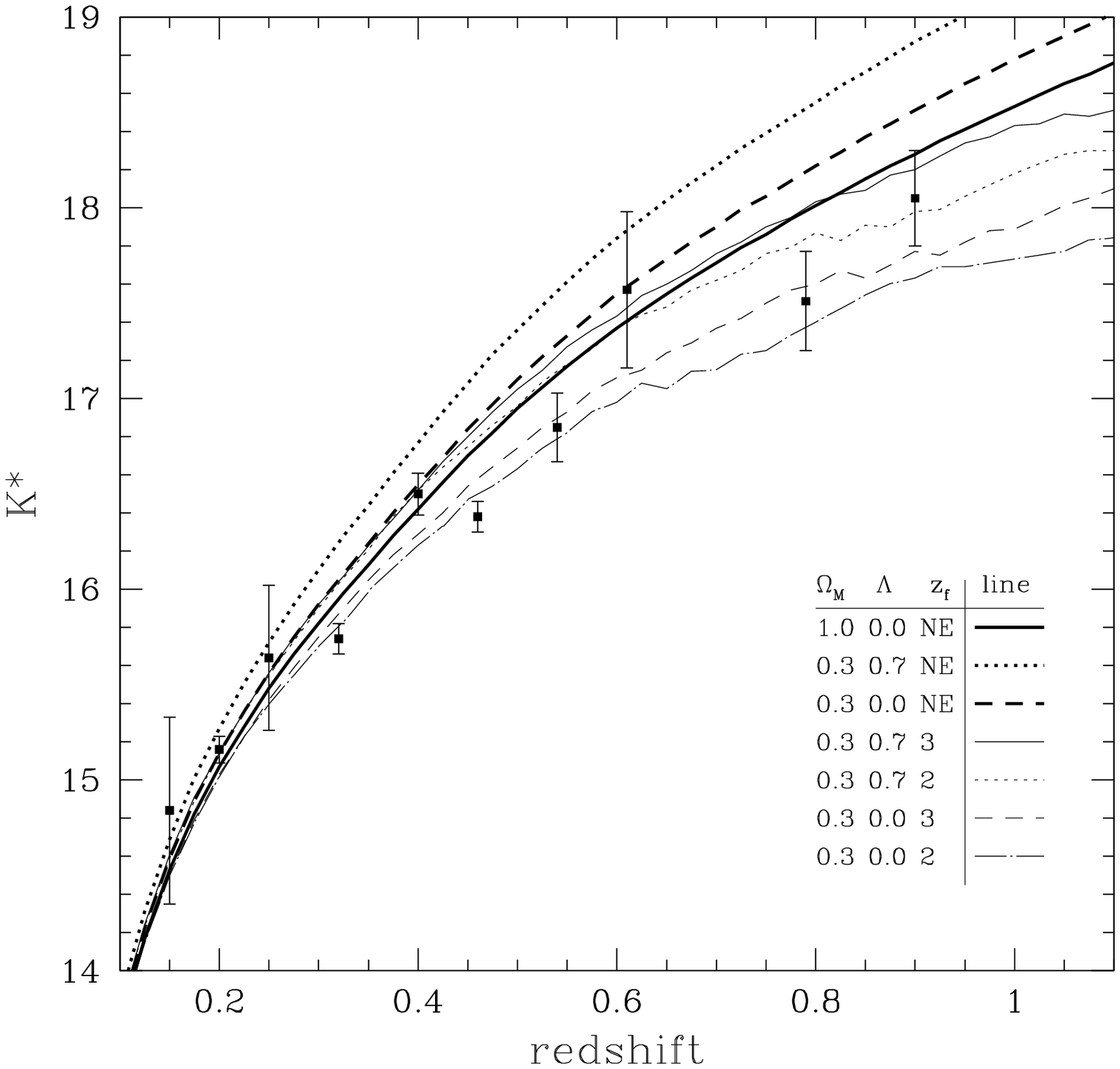}
\epsscale{1.0}
\end{figure}

\begin{figure}
\epsscale{1.1}
\caption{$K^*$ vs. $z$ for clusters with high--$L_X$ (solid squares) and low--$L_X$ 
(open triangles) compared with models for galaxy evolution. 
The models are the same as those as in Figure~8.}
\label{ksvszlx}
\plotone{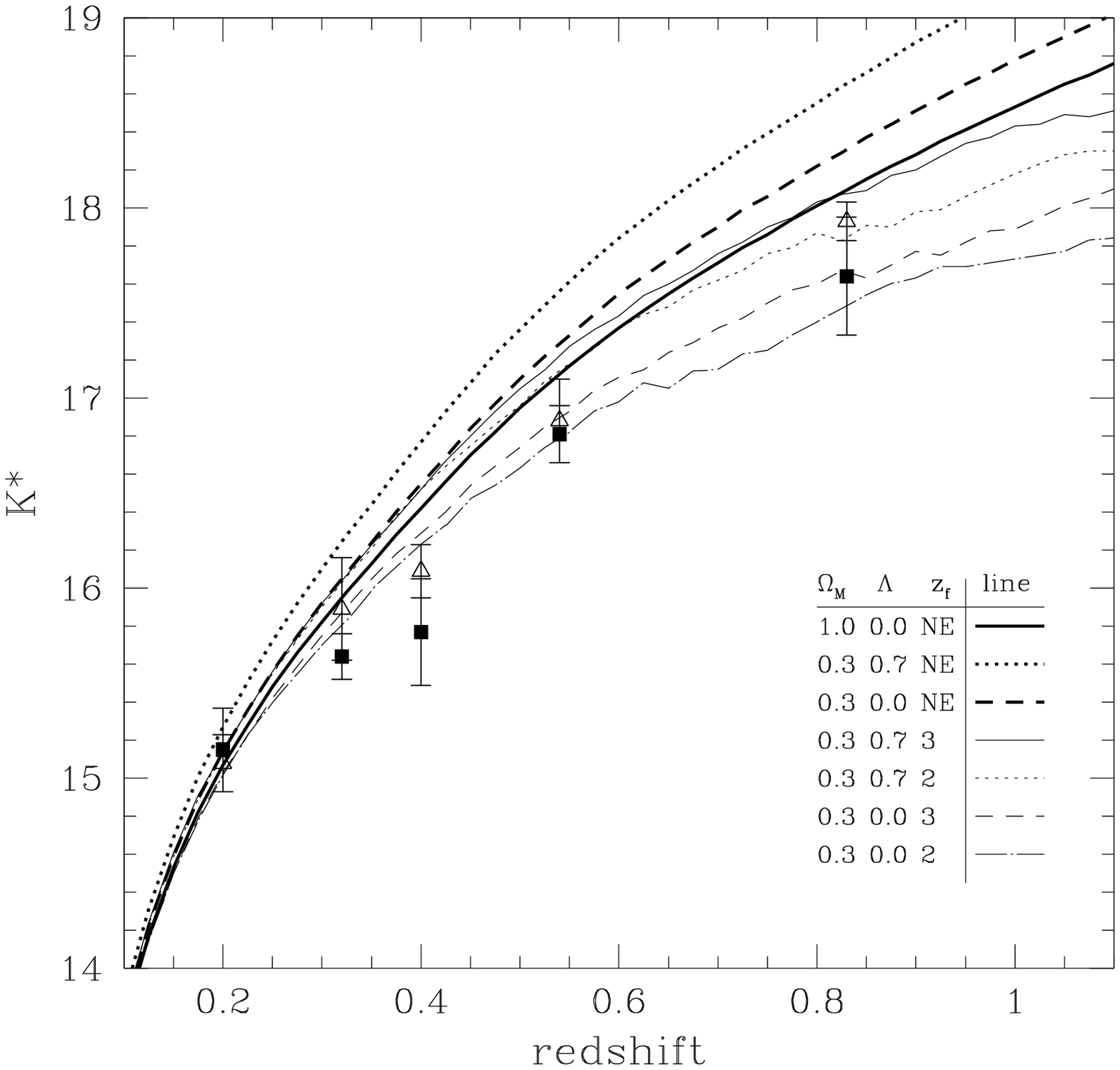}
\epsscale{1.0}
\end{figure}

\begin{figure}
\epsscale{1.1}
\caption{$K^*$ vs. $z$ for clusters in the rich (solid squares) and
poor (open triangles) subsamples compared with models for galaxy
evolution.  The models are the same as those as in Figure~8.}
\label{ksvszrp}
\plotone{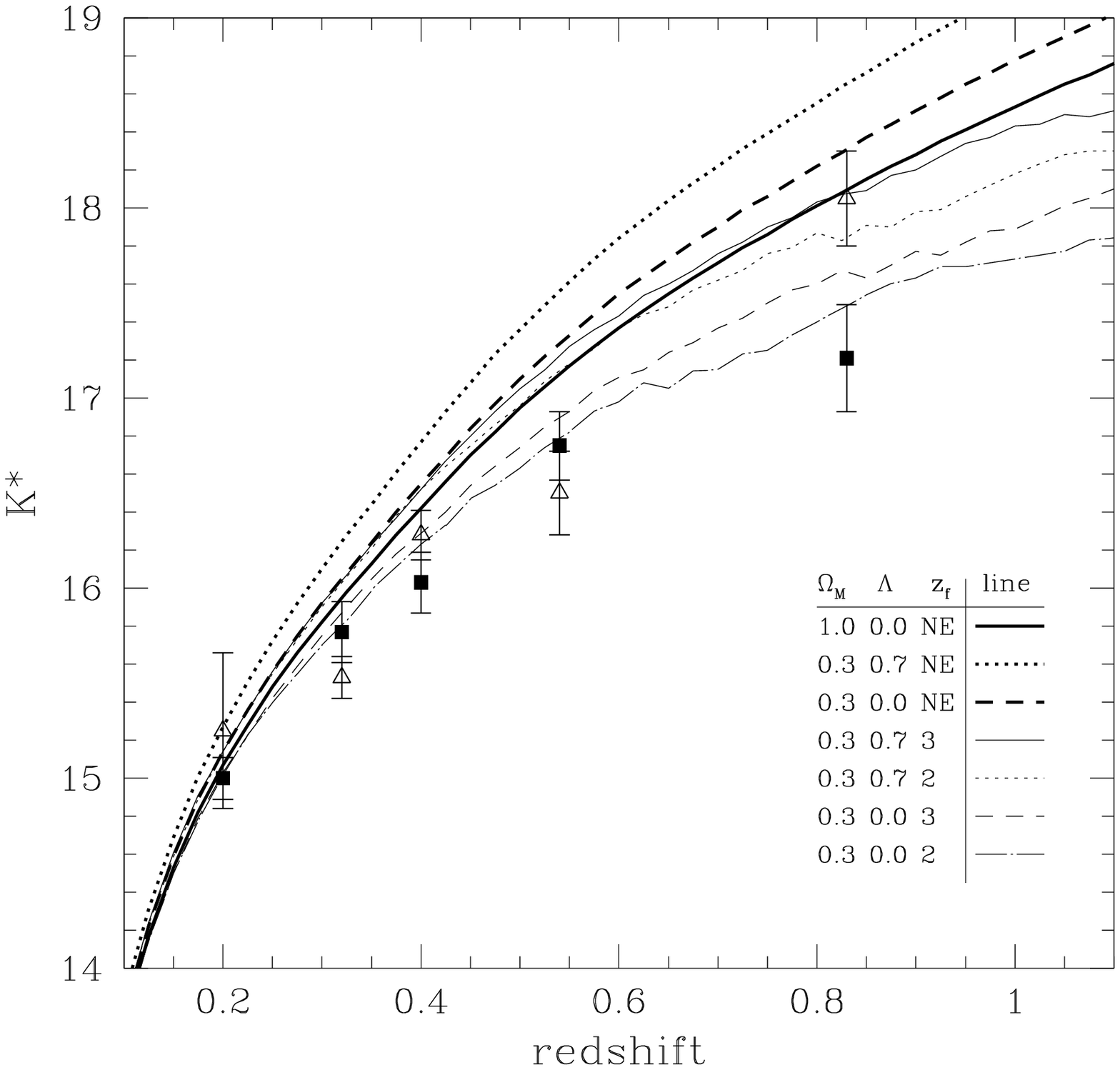}
\epsscale{1.0}
\end{figure}

\end{document}